\documentclass[letterpaper, aps, showpacs,onecolumn]{revtex4}
\usepackage{amsmath}
\usepackage{amssymb}
\usepackage{latexsym}
\usepackage[dvips]{graphics}
\usepackage{hyperref}

\begin{document}

\title{Trapping cold Atoms by Quantum Reflection}
\author{Alexander Jurisch and Jan-Michael Rost}
\affiliation{Max-Planck-Institut f\"ur Physik komplexer Systeme\\
N\"othnitzerstr. \!\!38, 01187 Dresden, Germany}
\begin{abstract}
We examine the properties of a quantum reflection trap when particle-interation is included. We explore the influence of the particle-interaction on the trapping for different regimes: repulsive particle-interaction and attractive particle-interactions in its stable and unstable limit. With variational techniques, we calculate the phase-diagram of the quantum reflection trap and determine the stable and unstable regimes of the system.
\end{abstract}
\maketitle
\section{Introduction}
Recent progress in experimentally controling BECs \cite{Pas1, Pas2} has shown, that not only single atoms, but condensed atomic clouds, Na in these cases, can be quantum reflected as a whole by atom-surface potentials when the kinetic energy of the incident cloud lies within the threshold region of the potential. Armed with these results, we suggested to trap ultra-cold atoms solely by quantum reflection in \cite{JurFri1}. The advantages of quantum reflection over other trapping mechanisms are obvious: stability of the reflection behaviour in the threshold region, no need for external auxiliary fields to form the trapping potential, because atom-surface interaction is delivered by nature free house. In \cite{JurFri1} we have shown, that the surviving particle density inside the trap as a function of time gives reasonable good results to be promising for future investigation. We have shown, that due to atom-surface interaction, an enhancement of the surviving particle density up to 50\% is achievable. The trapping times lie somewhat around 0.5 s, excluding mass factors of the atomic species used.

The model we proposed in \cite{JurFri1} relies solely on the threshold properties of an atom-surface potential. The trapping mechanism, quantum reflection, exploits the fact that cold atoms are reflected by an attractive potential tail without reaching a classical turning point. The mechanism of quantum reflection has enjoyed a growing interest, experimentally \cite{Shi, DruDeK, ObeKouShiFujShi, ObeTasShiShi, KouObe} and theoretically \cite{CotFriTro, FriJacMei, Jur}.

The equivalence between the true atom-surface potential and the step-potential at threshold comes from the fact, that both obey the same law of reflectivity, see, e.g., \cite{FriTro}. Very recently, Madronero and Friedrich, \cite{MadFri}, have shown that the dynamics of a wave-packet governed by a step-potential is indeed qualitatively similar to the dynamics of a wave-packet governed by a power-law atom-surface potential. Quantitatively, the differences are negligibly small and, as a matter of fact, the step-potential model even underestimates the results of the power-law potential. For quantum reflection, the use of a step-potential as a model for the atom-surface power-law potential has thus been justified once more.

Here, we will examine the properties of a spherical quantum reflection trap, which is the simplest choice of a trapping system when particle interaction is included. To model our potential in this case, we make the following assumptions on the behaviour of the atom-surface potential inside a sphere:
\begin{equation}
\lim_{r\,\rightarrow\,L}\,U(r)\,=\,-\,\frac{\hbar^{2}}{2\,m}\,\frac{\beta_{4}^{2}}{|r\,-\,L|^{4}},\quad
\lim_{r\,\rightarrow\,0}\,U(r)\,=\,0\quad,
\label{limittruepotential}\end{equation}
where $L$ is the radius of the sphere. The assumptions Eq. (\ref{limittruepotential}) are certainly true if the radius of the sphere is, by orders of magnitude, larger than the extension of the atomic wave-packet inside the sphere. If this condition is fullfilled, the inside wall of the sphere, for an atom close to it, must locally look like a plane, \cite{Wir}. In good agreement with these requirements, the atom-surface potential inside a sphere can be modeled as a step-potential
\begin{equation}
U(r) = -\,\frac{\hbar^{2}}{2\,m}\,b^{-2}\theta\left[r\,-\,L\right],\quad b\,=\,\beta_{4}\quad.
\label{potentialmodel}\end{equation}

\section{The Model of the Quantum Reflection Trap}
The system of interest is given by the Gross-Pitaevskii equation
\begin{equation}
i\,\hbar\,\frac{\partial}{\partial t}\Psi(r,t)\,=\,
\frac{\hbar^{2}}{2m}\,\left\{-\nabla^{2}\Psi(r,t)\,+\,\frac{2m}{\hbar^{2}}\,
U(r)\Psi(r,t)\,+\,
8\,\pi\,a_{{\rm{int}}}\,\left|\Psi(r,t)\right|^{2}\,\Psi(r,t)\right\}\quad, 
\label{systemgen}\end{equation}
where the particle density is normalized to unity. The scale of the system is naturally given by the spatial extension of the trap, the radius $L$ of the sphere. The scaling variables are thus given by \cite{JurFri1},
\begin{equation}
x\,=\,\frac{r}{L},\quad \sigma\,=\,\frac{L}{\beta_{4}},\quad \tau\,=\,\frac{t\,\hbar}{2\,m\,L^{2}},\quad \Psi\,\rightarrow L^{-\frac{3}{2}}\Psi\quad.
\label{scalingvariables}\end{equation}
The last term in Eq. (\ref{scalingvariables}) restores normalization to unity.
The atom-surface potential Eqs. (\ref{limittruepotential}, \ref{potentialmodel}) shows only a dependence in the radial direction, so the angles can be separated off. Assuming the system to be in an s-wave state, the scaled wave-function is
\begin{equation}
\Psi(x,\tau)=\,\frac{\psi_{{\rm{rad}}}(x,\tau)}{x}\,Y_{00}(\vartheta,\varphi),\quad Y_{00}(\vartheta,\varphi)\,=\,\frac{1}{\sqrt{4\,\pi}}\quad.
\label{3dimwavefunction}\end{equation}
With Eq. (\ref{3dimwavefunction}), the angular parts can easily be integrated out which yields for the interaction energy
\begin{equation}
\frac{8\,\pi\,a_{{\rm{int}}}}{L}\,\int_{-\infty}^{\infty}d^{3}x\,\left|\Psi(x,\tau)\right|^{4}\,=\,
\frac{8\,\pi\,a_{{\rm{int}}}}{4\,\pi\,L}\int_{0}^{\infty}dx\,\frac{\left|\psi_{{\rm{rad}}}(x,\tau)\right|^{4}}{x^{2}}\quad.
\label{radialpartinteraction}\end{equation}
The radial coupling-constant is then given by
\begin{equation}
\gamma_{{\rm{rad}}}\,=\,\frac{8\,\pi\,a_{{\rm{int}}}}{4\,\pi\,L}\,=\,\frac{2\,a_{{\rm{int}}}}{L}\quad.
\label{radialcoupling}\end{equation}

The radial Gross-Pitaevskii equation describing our quantum reflection trap in scaled form thus finally reads
\begin{equation}
i\,\frac{\partial}{\partial\,\tau}\,\psi(x,\tau)\,=\,-\frac{\partial^{2}}{\partial x^{2}}\psi(x,t)\,-\,\sigma^{2}\theta\left[x-1\right]\,\psi(x,\tau)\,+\,\gamma\frac{\left|\psi(x,\tau)\right|^2}{x^{2}}\,\psi(x,\tau)\quad.
\label{GP}\end{equation}
To simplify our notation, we have dropped the subscript of the radial wave-function and the interaction-strength $\gamma$. The step function on the right hand side in Eq. (\ref{GP}) is again the model of the atom-surface interaction.

In table (\ref{couplingconstants}) we have listed some values of the radial coupling constant for the alkali atoms typically used in BEC experiments.
\begin{table}[!h]
\begin{tabular}{c||c c|| c ||c c c}
\hline
\hline
 & $\sigma$\,=\,$L/\beta_{4}$ & $\beta_{4}$\,[a.u.] & $L$\,[a.u.] &
 $\gamma$ & $a^{{\rm{singlet}}}_{{\rm{int}}}$\,[a.u.] & $a^{{\rm{triplet}}}_{{\rm{int}}}$\,[a.u.]\\
\hline
${}^{6}$Li & 54.25  & $8.239 \times 10^{3}$ & $4.47 \times 10^{5}$ & $-\,9.66\,\times\,10^{-3}\,\pm\,7.95\,\times\,10^{-4}$ & -- & $-\,2160\,\pm\,250$\\
\hline
${}^{23}$Na & 30 & $1.494 \times 10^{4}$ & $4.47 \times 10^{5}$ & $2.92 \times 10^{-4} \pm\, 3.97 \times 10^{-6}$ & -- & $65.3\,\pm\,0.9$\\
\hline 
${}^{85}$Rb & 11 & $4.033 \times 10^{4}$ & $4.47 \times 10^{5}$ & $0.01_{-0.0008}^{+0.002}$ & $2400_{-350}^{+600}$ & -- \\
\hline
\hline
\end{tabular}
\caption{Comparison of the scaled potential strength $\sigma$ and the coupling-constant $\gamma$ for some alkali atoms. The radial extension of the trap is allover $L\,=\,4.47\,\times\,10^{5}$\,[a.u.]\,. The potential data was taken from \cite{FriJacMei}, the data for the scattering lengths were taken from \cite{AbrMcAGerHulCotDal} for Li, from \cite{AbeVer} for Na, and form \cite{RobClaBurGreCorWie, CorClaRobCorWie} for Rb\,.}
\label{couplingconstants}\end{table}
Table (\ref{couplingconstants}) shows that the natural influence of the coupling due to the particle-interaction is rather small, such that the results published in \cite{JurFri1}, where trapping was considered neglecting particle-interaction, are confirmed to apply for cases where the particle-interaction strength is not tuned. On the other hand, a tuning of the coupling constant by the help of a Feshbach-resonance delivers a wider range for $\gamma$, which we now will examine theoretically.

\section{Wave-packet dynamics with particle interaction}
To explore the dynamical properties of an atomic wave-packet with particle-interaction inside a quantum reflection trap, we solve Eq. (\ref{GP}) with the initial condition
\begin{equation}
\psi(x,\tau\,=\,0)\,=\,N\,x\,\exp\left[-\,a\,x\right]\theta\left[1\,-\,x\right]\quad,
\label{initialwavepacket}\end{equation}
where $N$ is the normalization constant and $a$ is the diffuseness of the wave-packet. The wave-packet Eq. (\ref{initialwavepacket}) provides the simplest possible choice for an initial state supporting the main parts of particle density around $x\,=\,0$, \cite{JurFri1}. To be more explicit, we take the example of Na-atoms, which have been used in the recent quantum reflection experiments with BECs \cite{Pas1, Pas2}. A diffuseness of $a\,=\,5$, together with the atomic mass of sodium and a trap radius of $L\,=\,4.5\,\times\,10^{5}$\,[a.u.] gives an initial kinetic energy $E_{\rm{kin}} = 1.5\,\times\,10^{-15}$\,[a.u.] that corresponds to temperatures of approximatly  1\,nK.

The observables of our examinations are the surviving particle density $\rho_{\rm{S}}(\tau)$ inside the trap, and the scaled energy $\mathcal{E}(\tau)$ of the system, as functions of time:
\begin{equation}
\rho_{\rm{S}}(\tau)\,=\,\int_{0}^{1}\,dx\,\left|\psi(x,\tau)\right|^{2},\quad 
\mathcal{E}(\tau)\,=\,i\int_{0}^{1}\,dx\,\psi^{*}(x,\tau)\partial_{\tau}\psi(x,\tau)\quad.
\end{equation}

Numerically, we solve our differential equation (\ref{GP}) by using the Crank-Nicholson algorithm and employ absorbing boundary conditions, see \cite{NeuBae}, to simulate outgoing waves, that have left the spatial region of the quantum reflection trap.

\subsection{Dynamics with repulsive particle-interaction}
In cases where the coupling constant $\gamma$ is larger than zero, the particle-interaction is repulsive. The additional positive energy due to the self-interaction increases the total energy of the system. An increase of the systems's total energy slightly accelerates the decay of the surviving particle fraction inside the trap, because even low-energetic components of the wave-packet gain additional energy, which facilitates the escape from the trap.
\begin{figure}[h!]\centering
\rotatebox{-90.0}{\scalebox{0.4}{\includegraphics{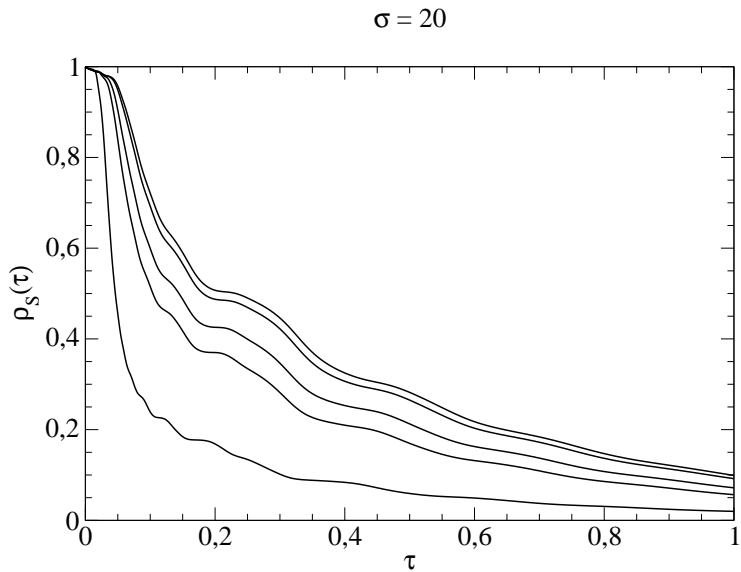}}}
\caption{\footnotesize{Surviving partile densities for $\sigma\,=\,20$ and values of $\gamma\,=\,0, 0.1, 0.5, 1.0, 5.0$ from top to bottom. The presence of the repulsive particle interaction reduces the trapping effect of the quantum reflection by its additional contribution to the total energy of the system.}}
\label{Densitysigma20}\end{figure}
As an example we have shown in Fig. (\ref{Densitysigma20}) the surviving particle density $\rho_{S}(\tau)$ for a potential strength of $\sigma\,=\,20$ and several values of $\gamma\,\geq\,0$. The shape of $\rho_{S}(\tau)$ with finite $\gamma$ is similar to the shape of the curve with $\gamma\,=\,0$, see \cite{JurFri1}, but all surviving particle densities for finite $\gamma$ are smaller than for $\gamma\,=\,0$. The results of figure (\ref{Densitysigma20}) confirm, that the only influence of the particle-interaction on quantum reflection is indeed given by a slight acceleration of the decay of the surviving particle density inside the trap, that stems from an enhanced total energy. Other effects due to particle-interaction should modify the shape of the densities as functions of time, but the typical plateaus, the genuine pattern of quantum reflection already found in \cite{JurFri1} are conserved.
\begin{table}[!h]
\begin{tabular}{c||c  c  c  c  c}
\hline
\hline
$\rho_{S}(\tau = 1)$,\, $\gamma\,=$  & 0 & 0.1 & 0.5 & 1.0 & 5.0 \\
\hline
$\sigma\,=\,20$ & 0.11 & 0.10 & 0.071  & 0.056  & 0.019 \\
\hline
$\sigma\,=\,30$ & 0.19 & 0.17 & 0.14  & 0.11  & 0.043 \\
\hline
$\sigma\,=\,40$ & 0.27 & 0.24 & 0.20  & 0.16  & 0.06 \\
\hline
$\sigma\,=\,50$ & 0.33 & 0.31 & 0.25 & 0.19 & 0.09 \\
\hline
\hline
\end{tabular}
\caption{\footnotesize{Listed are the surviving particle densities for several values of $\sigma$ and $\gamma$ for repulsive interaction.}}
\label{ValuesSigma20}\end{table} 
The values of table (\ref{ValuesSigma20}) show that the surviving particle fractions after the scaled time $\tau\,=\,1$ - approximatly one half of a second for the alkali atoms, see Eq. (\ref{scalingvariables}) - depend only weakly on the self-interaction when $\gamma\,<\,1$ . Especially for higher values of $\sigma$, the surviving particle densities for $\tau\,=\,1$ and $\gamma\,\leq\,1$ give still good results for trapping. Compared with a freely spreading wave-packet, which has a value $\rho_{S}(\tau\,=\,1)\,=\,0.005$, \cite{JurFri1}, the surviving particle densities for $\gamma\,=\,1$ are still enhanced by factors 10 to 40.
For $\gamma\,>\,1$, the particle density inside the trap starts to decay rapidly. For this regime, the mechanism of quantum reflection, although still working, is not strong enough for an effective trapping of atomic wave-packets, because the total energy of the system is too strongly increased by the interaction potential.
\begin{figure}[h!]\centering
\rotatebox{-90.0}{\scalebox{0.4}{\includegraphics{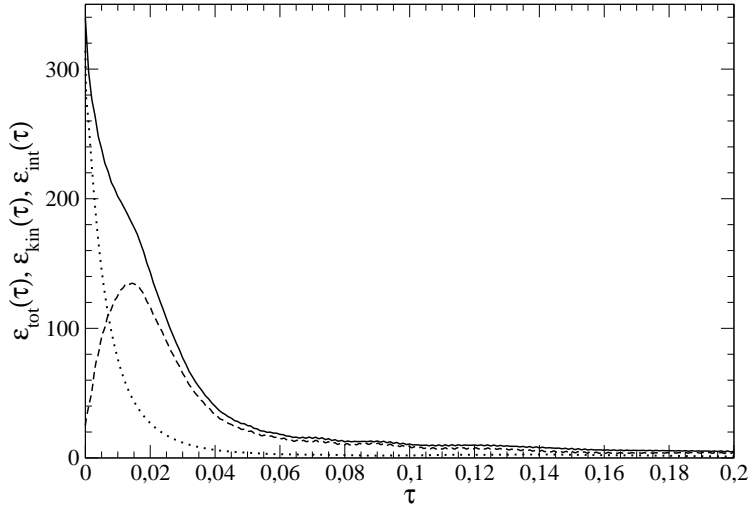}}}
\caption{\footnotesize{Shown are the total energy of the system $\mathcal{E}_{\rm{tot}}(\tau)$ (full line), the kinetic energy $\mathcal{E}_{\rm{kin}}(\tau)$ (dashed line) and the repulsive particle-interaction $\mathcal{E}_{\rm{int}}(\tau)$ (dotted line) for $\sigma\,=\,40$ and $\gamma\,=\,5$.}}
\label{Energysigma20g5}\end{figure}
As seen from Fig. (\ref{Energysigma20g5}), the strong repulsive interaction dominates the system at the beginning of the time-evolution, but as the atomic wave-packet evolves in time, there is a continous current density of high-energy components beyond the edge of the step and a change of the shape of the wave-packet due to its motion on the step. The change of the shape influences the self-interacting potential and continously reduces its influence on the dynamics. But as the interaction energy decreases, it is transformed into kinetic energy. This leads to a faster motion and thus to a reduction of the effect of quantum reflection when large fractions of high-energetic components reach the edge of the step. Already after a time $\tau \approx 0.1$ the system has strongly cooled down, but along with a strong loss of particle density. In short: the larger the kinetic energy of the wave-packet, the weaker is the effect of the quantum reflection.

Therefore, we may conclude that a scaled particle-interaction strengths $\gamma\,\leq\,1$ effectively traps atoms. This is particulary true for large values of $\sigma$, which requires a weak potential strength $\beta_{4}$, see Eqs. (\ref{limittruepotential}, \ref{potentialmodel}).

\subsection{Dynamics with attractive particle-interaction}
\begin{figure}[h!]\centering
\rotatebox{-90.0}{\scalebox{0.4}{\includegraphics{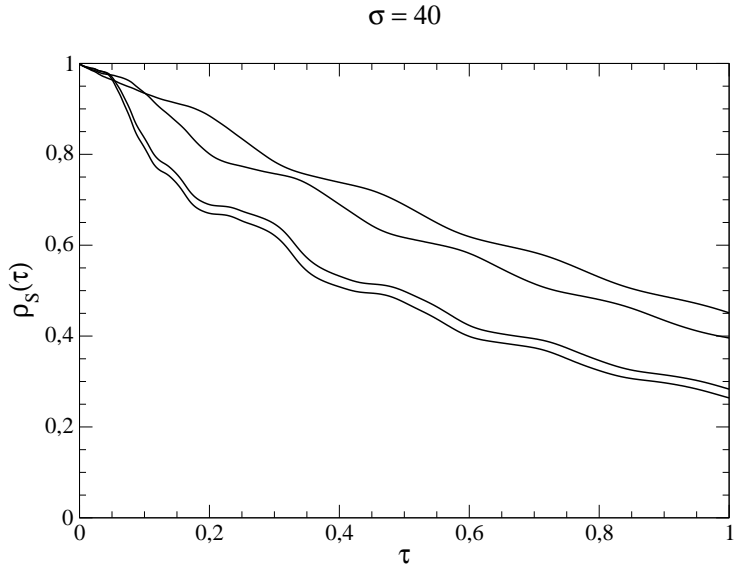}}}
\caption{\footnotesize{Surviving partile densities for $\sigma\,=\,40$ and values of $\gamma\,=\,0, -0.1, -0.5, -0.62$ from bottom to top. The presence of the attractive particle-interaction increases the trapping effect of the quantum reflection.}}
\label{Dichtesigma40}\end{figure}

In cases where the coupling constant $\gamma$ is negative, the particle-interaction is attractive. The additional, negative energy due to the self-interaction reduces the total energy of the system. As a consequence, the decay of the surviving particle fraction inside the trap is remarkably decelerated. The presence of attractive particle-interaction leads to a self-trapping of all lower momentum components which therefore are much more unlikely to even reach the edge of the trapping potential. 
In figure (\ref{Dichtesigma40}) we show surviving particle densities for different strength $\gamma$ of the self-interaction. For small interaction strength, $\gamma\,=\,-0.1$, the dynamics is similar to the non-interacting case, $\gamma\,=\,0$, while for increasing interaction towards the critical interaction strength for collapse, we observe that the plateau structure becomes significantly washed out, $\gamma\,=\,-0.5$, $\gamma\,=\,-0.62$.

Numerically we have determined the critical value $\gamma_{c}\,=\,-0.627$. For attractive interactions stronger than $\gamma_{c}$, the wave-packet dynamics will undergo a collapse. As will be discussed below, the value of $\gamma_{c}$ is universal for of the quantum reflection trap and a wider class of systems, which have no classical longitudinal (radial) confinement.

The drastic change of the surviving particle density in the vincity of $\gamma_{c}$ is reminiscent of the well-known phenomenon of a critical slowing down near the point of a phase-transition. 
\begin{table}[!h]
\begin{tabular}{c||c  c  c  c  c}
\hline
\hline
$\rho_{S}(\tau = 1)$,\, $\gamma\,=$  & 0 & -0.1 & -0.5 & -0.6 & -0.62 \\
\hline
$\sigma\,=\,20$ & 0.11 & 0.13 & 0.17 & 0.20  & 0.21 \\
\hline
$\sigma\,=\,30$ & 0.19 & 0.21 & 0.30 & 0.34  & 0.35 \\
\hline
$\sigma\,=\,40$ & 0.27 & 0.28 & 0.40 & 0.44  & 0.45 \\
\hline
$\sigma\,=\,50$ & 0.33 & 0.35 & 0.47 & 0.51 & 0.52 \\ 
\hline
\hline
\end{tabular}
\caption{\footnotesize{Listed are the surviving particle densities for several values of $\sigma$ and $\gamma$ for attractive interaction. The reference time $\tau\,=\,1$ corresponds to around half a second for the here considered alkali species, see eq. (\ref{scalingvariables}). The surviving particle densities show that trapping by quantum reflection gives truely promising results in this regime.}}
\label{ValuesSigma40}\end{table}
The values of table (\ref{ValuesSigma40}) show that the surviving particle fractions $\rho_{{\rm{S}}}(\tau\,=\,1)$ can achieve a strong enhancement when the particle-interaction strength $\gamma$ lies in the regime of $0\,\geq\,\gamma\,\geq\,-0.627$. Compared to the case of $\gamma\,=\,0$, the enhancement factor ranges from 1.6 to 2.0, but compared to the freely decaying wave-packet, $\rho_{{\rm{S}}}(\tau\,=\,1)\,=\,0.005$, \cite{JurFri1}, the trapping mechanism achieves enhancement factors from 42 for $\sigma\,=\,20$ to 104 for $\sigma\,=\,50$. Clearly, negative self-interaction, compared to the noninteracting case, significantly enhances the efficiency of the trap.
\begin{figure}[h!]\centering
\rotatebox{-90.0}{\scalebox{0.4}{\includegraphics{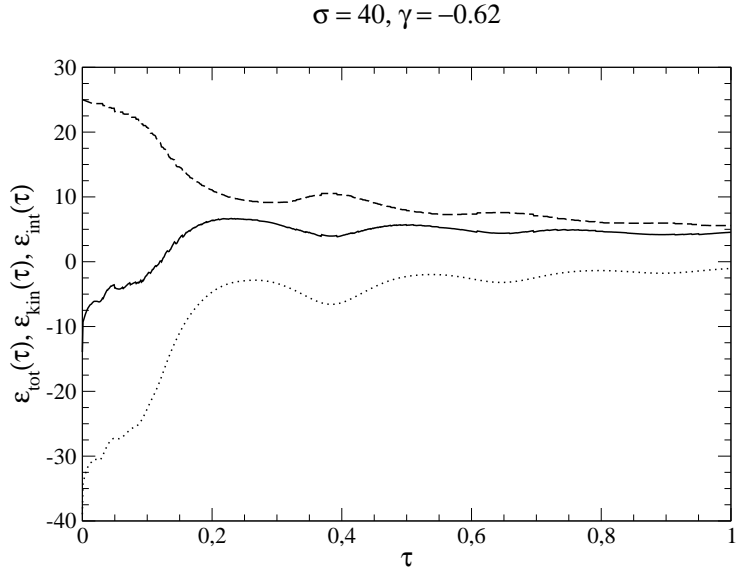}}}
\caption{\footnotesize{We have drawn the total energy of the system $\mathcal{E}_{\rm{tot}}(\tau)$ (full line), the kinetic energy $\mathcal{E}_{\rm{kin}}(\tau)$ (dashed line) and the repulsive particle-interaction $\mathcal{E}_{\rm{int}}(\tau)$ (dotted line) for $\sigma\,=\,40$ and $\gamma\,=\,-0.62$. It can clearly be seen how the energetic components of the system balance each other. The dynamical equilibrium between the kinetic energy and the interaction-energy stabilizes the system.}}
\label{Energysigma40}\end{figure}
Figure (\ref{Energysigma40}) illustrates the effect of trapping under self-interaction on the energies of the system as functions of time. At the beginning of the time-evolution, the total energy of the system is negative, and the particle-interaction dominates the system. The process of stabilization of the wave-packet by the self-trapping effect establishes a state of dynamical equilibrium between the kinetic energy and the interaction energy. When the total energy of the system has become positive, the wave-packet is no more in danger to collapse and has entered the regime were the kinetic energy, together with the quantum reflection, govern the behaviour of the system. However, the kinetic energy of the wave-packet has been reduced by the stabilization process, rendering quantum reflection even more effective. Together with figure (\ref{Dichtesigma40}) the behaviour of the energy clearly explains the trapping in this regime.

\subsection{Dynamical collapse for strong attractive interaction}
\begin{figure}[h!]\centering
\rotatebox{-90.0}{\scalebox{0.4}{\includegraphics{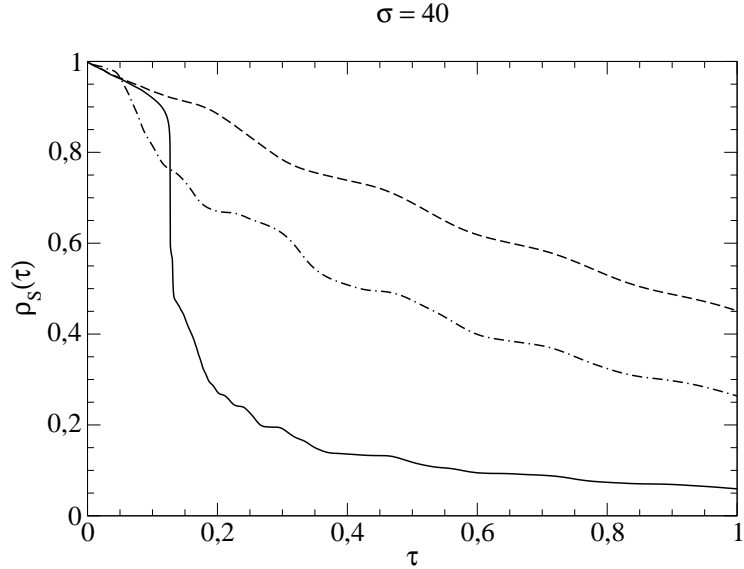}}}
\caption{\footnotesize{Surviving partile densities for $\sigma\,=\,40$ and values of $\gamma\,=\,-0.63$ (full line), $\gamma_{c}\,=\,-0.627$ (dashed line) and $\gamma\,=\,0$ (dot-dashed line) for comparison. The collapse of the wave-packet is indicated by the sudden fall-off of the surviving particle density.}}
\label{Dichtesigma40gminus063}\end{figure}

\begin{figure}[h!]\centering
\rotatebox{-90.0}{\scalebox{0.4}{\includegraphics{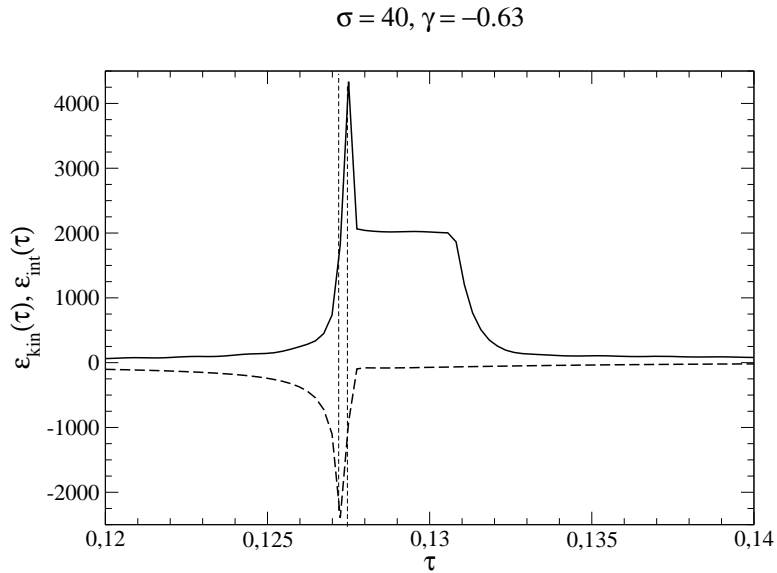}}}
\caption{\footnotesize{Shown are the kinetic energy $\mathcal{E}_{\rm{kin}}(\tau)$ (full line) and the repulsive particle-interaction $\mathcal{E}_{\rm{int}}(\tau)$ (dashed line) for $\sigma\,=\,40$ and $\gamma\,=\,-0.63$. The peak of the kinetic energy is responsible for the break-down of the trapping mechanism and the strong and sudden loss of particle density, see Fig. (\ref{Dichtesigma40gminus063}).}}
\label{Energysigma40gminus063}\end{figure}

The dynamics of a wave-packet in the quantum reflection trap becomes unstable against collapse, when the interaction strength $\gamma$ becomes smaller than the critical value $\gamma_{c}\,=\,-0.627$. As the wave-packet evolves in time, the influence of the attractive particle-interaction grows strongly, leading to a strong localization of the wave-packet. Strong localization goes along with high momenta, which lead to the destruction of the coherence of the wave-packet. When the wave-packet undergoes a complete disruption, it collapses. See the full curve in figure (\ref{Dichtesigma40gminus063}). The collapse goes along with a strong loss of particle density, that takes place during an extremely short interval of time. After the collapse of the wave-packet the attractive self-interaction is also destroyed and the effect of quantum reflection again dominates the behaviour of the system. The effect of quantum reflection manifests itself by stopping the evaporation and forcing the wave-packet to reshape. Thus, the remaining but strongly reduced density in the trap decays regularly as the system continues to evolve in time.
Figure (\ref{Energysigma40gminus063}) shows the behaviour of both energies. To help the readers' eyes in separating the critical peaks of both energies, we have drawn two vertical lines. It can be clearly seen how the negative interaction energy grows to strong negative values during the time-evolution of the system. The kinetic energy, on the positive half of the ordinate also grows during collapse in the attempt to achieve a dynamical equilibrium that stabilizes the system. However, by this peak in the kinetic energy the mechanism of quantum reflection becomes meaningless and a large fraction of particle density simply evaporates. Even after its peak the kinetic energy of the system is large enough to continuously drive a considerable fraction of particle density to the outside, such that the system experiences a self-cooling and the kinetic energy falls off again. When this has happened the effect of quantum reflection regains the control over the dynamics. Note, that the particle-interaction, due to the strong loss of particle density, remains negligible.

\section{Variational stability analysis}
The stability of our quantum reflection trap can be accessed by variational techniques. A very elaborated and general variational approach to cold atom systems was given in \cite{PerLewCirZol}, where the critical interaction-strength $\gamma_{c}$ for an anisotropic harmonically trapped 3D BEC-system in the absence of a longitudinal confinement was determined to $\gamma_{c}\,=\,-0.6204$. The same problem is considered in \cite{CarCas} numerically, where $\gamma_{c}\,=\,-0.627$ was found. For our system, we have found the same critical value $\gamma_{c}\,=\,-0.627$, despite a longitudinal (radial) confinement due to the atom-surface potential. The difference can be explained by the fact that the atom-surface potential is confining with respect to quantum reflection, whereas a harmonic oscillator potential is confining due to classical reflection at the potential surface. When a longitudinal confinement is present in a harmonic oscillator system, the critical value is reduced to $\gamma_{c}^{{\rm{ho}}}\,=\,-0.57$, \cite{RupHolBurEdw}. The absence of a (classical) longitudinal confinement due to a potential surface, along with the presence of a quantum confinement provided by atom-surface potentials thus enables the storage of a higher number of atoms for attractive particle-interaction on macroscopic time-scales.

To analyze the stability of our system, we will refer to a much simpler variational technique than the one suggested in \cite{PerLewCirZol}, see e.g. \cite{DalGioPitStr}, where the authors restrict themselves to a simple gaussian trial-function. Our choice for a radial trial-function defined on the whole space is
\begin{equation}
\phi(x)\,=\,N\,x\,\exp\left[-\frac{x^{2}}{2\,\alpha^{2}}\right]\quad,
\label{gausstrial}\end{equation}
where $N$ is the normalization and $\alpha$ is the width.
With Eq. (\ref{gausstrial}) we obtain a parametrized energy
\begin{equation}
\mathcal{H}(\alpha, \sigma, \gamma)\,=\,\left<\phi\left|\hat{H}\right|\phi\right>\,=\,
\frac{3}{2\,\alpha^{2}}\,-
\,\sigma^{2}\left(\frac{2}{\sqrt{\pi}}\exp\left[-\,\alpha^{-2}\right]\,-\,\alpha\,{\rm{erfc}}\left[\alpha^{-1}\right]\right)\,+\,\gamma\,\sqrt{\frac{2}{\pi\,\alpha^{6}}}\quad.
\label{paramenergy}\end{equation}
With Eq. (\ref{paramenergy}) we can estabilish a relation between the particle-interaction $\gamma$ and the width of the gaussian state $\alpha$ by demanding that the variation of $\mathcal{H}$ with respect to $\alpha$ vanishes,
\begin{equation}
\delta\,\mathcal{H}\left(\alpha,\sigma,\gamma(\alpha)\right)\,=\,0\quad.
\label{variationH}\end{equation}
In Eq. (\ref{variationH}) we have kept $\sigma$ as a parameter. Solving Eq. (\ref{variationH}) for $\gamma(\alpha,\sigma)$ leads to
\begin{equation}
\gamma(\alpha,\sigma)\,=\, -\,\alpha^{2}\,+\,\frac{\alpha}{\sqrt{2}}\left[2\,\alpha\,\sigma^{2}\,\exp\left[-\alpha^{-2}\right]\,+\,
\alpha^{2}\,\sigma^{2}\,\left(1\,-\,{\rm{erf}}\left[\alpha^{-1}\right]\right)\,-\,\frac{3}{2}\,\sqrt{\pi}\right]\quad.
\label{runningcoupling}\end{equation}
Equation (\ref{runningcoupling}) defines a running coupling constant as a function of the initial width $\alpha$ of the wave-packet and the strength of the atom-surface potential $\sigma$. The coupling constant depends implicitly on the initial kinetic energy of the wave-packet, which, as we have seen above, is crucial for the efficiency of the trapping mechanism. With $a\,=\,5$ in Eq.(\ref{initialwavepacket}) we obtain $\alpha\,\approx\,0.31$ for Eq. (\ref{gausstrial}).

\begin{figure}[h!]\centering
\rotatebox{-90.0}{\scalebox{0.5}{\includegraphics{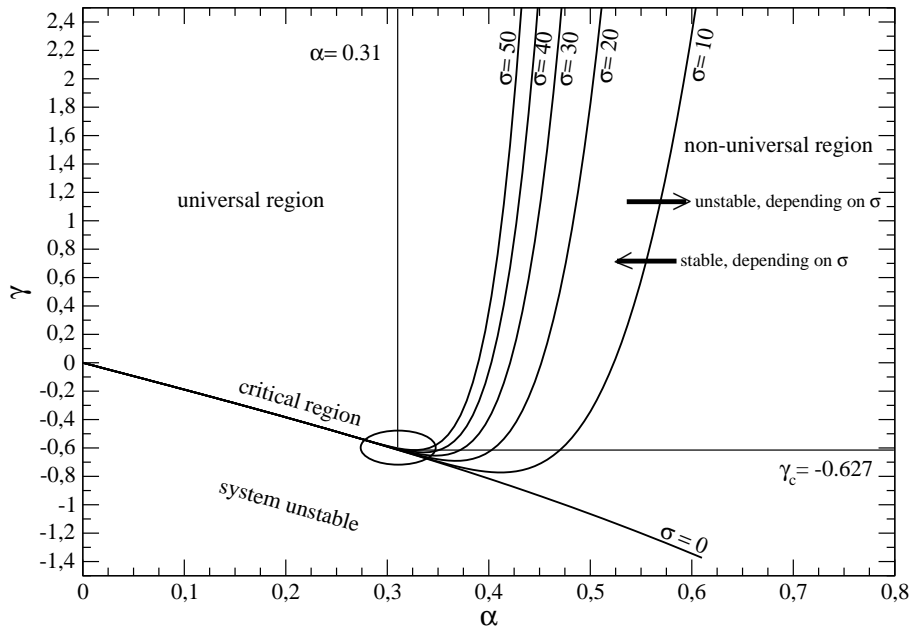}}}
\caption{\footnotesize{Phase-diagram of the quantum reflection trap obtained by a simple gaussian variational approach. The vertical line $\alpha\,=\,0.31\,\leftrightarrow\,a\,=\,5$ marks our initial data, the horizontal line $\gamma_{c}\,=\,-0.627$ marks our numerical result for the critical interaction-strength. The region encircled by the ellipse is the critical region of the system where transitions between universal and non-universal behaviour, as well as transitions between stable and unstable states of the condensate occur. The region below $\sigma\,=\,0$ supports only unstable states. The arrows drawn on the phase-line of $\sigma\,=\,10$ indicate exemplarily for any other value of $\sigma$, that a stable state for a certain value of $\sigma$ becomes unstable when its parametric set ($\alpha, \gamma$) is crossing the corresponding phase-line from left to right and vice versa.}}
\label{runningcouplingplot}\end{figure}

From Eq. (\ref{runningcoupling}), we obtain the phase-diagram of the quantum reflection trap. From the phase-diagram, Fig. (\ref{runningcouplingplot}), which we have drawn for all values of $\sigma$ considered above, the regions of stability can be read off easily and the value $\gamma_{c}\,=\,-0.627$ emerges as universal property of the quantum reflection trap. The horizontal and vertical lines mark our numerically determined value $\gamma_{c}\,=\,-0.627$ and our initial data $a\,=\,5$, respectively. 

As can be read off from figure (\ref{runningcouplingplot}), the simple gaussian trial-function delivers a critical region close to the exact value, such that we expect only small corrections from the general method of \cite{PerLewCirZol}. Also, our initial choice for $a\,=\,5$, which matches the experimental conditions due to \cite{Pas1, Pas2} is justified as optimal choice. The universal region of the system is located left from the vertical line denoting $a\,=\,5\,\leftrightarrow\,\alpha\,=\,0.31$, where no diversification of $\gamma$ due to the potential strength $\sigma$ occurs. By increasing $a$, which means reducing $\alpha$, the storage capacity of the system is reduced, because the critical value $\gamma_{c}$ is not accessible anymore. Vice versa the same occurs, because by decreasing $a$, which means increasing $\alpha$, the region of universality is left and the stability of the system is only granted along the phase-lines described by the strength of the atom-surface potential $\sigma$, requiring a higher value for $\gamma$, which also reduces the storage capacity of the system for attractive particle-interaction.

Lastly, Fig. (\ref{runningcouplingplot}), together with our numerical results from above proves, that the system shows not much sensitivity to the shape of the initial state, which neccessarily includes some arbitrariness in theoretical considerartions. 

\section{Summary and Conclusion}
We have investigated the possibility of trapping cold atoms solely by the mechanism of quantum reflection. Our investigations have been carried out with the simplest possible model of a quantum reflection trap, a sphere. Typically the radius $L$ of such a trapping sphere is of the order $10^{5}$\,[a.u.]. The quantum reflection is mainly controlled by the strength of the atom-surface potential $\sigma\,=\,L/\beta_{4}$. The larger $\sigma$, the better does quantum reflection work. The strength parameter $\sigma$ can be controlled either by the spatial extension of the trap, $L$, or by the atomic interaction parameter $\beta_{4}$. Small values of $\beta_{4}$ can be achieved, e.g., by using dielectric instead of perfectly conducting surfaces, see \cite{Jur, FriTro} and references therein. However, instead of controlling $\sigma$ by surface-engineering, it may be much easier to control the value of $\gamma$ by a Feshbach-resonance.

The inclusion of repulsive particle interaction, $\gamma\,>\,0$ depletes the surviving particle density inside the trap in comparison to a system evolving without particle interaction. As long as the interaction strength $\gamma$ is smaller than unity, the depletion of the surviving particle density due to the repulsive interaction can be compensated by increasing the atom-surface interaction strength $\sigma$, see table (\ref{ValuesSigma20}). 

The best results for trapping atoms by quantum reflection are achieved, when the particle interaction is attractive. The system remains stable when $\gamma$ lies in the range of $0\,>\,\gamma\,\geq\,\gamma_{c}$. The critical value $\gamma_{c}\,=\,-0.627$ is a universal property of our quantum reflection trap. This value of $\gamma_{c}$ makes clear that cold atom systems confined by an atom-surface potential belong to a wider class of systems with classical longitudinal freedom. Classical means, that no confinement due to a potential surface is present. The mechanism of quantum reflection acting in the case of atom-surface potentials establishes a quantum confinement. The quantum confinement has the advantage that more particles can be stored as in the case of a classical confinement, where the critical value is $\gamma_{c}^{{\rm{ho}}}\,=\,-0.57$.

Approaching the critical value $\gamma_{c}$ from above, the typical plateau structure of $\rho_{{\rm{S}}}(\tau)$ is washed out into slowly varying density oscillations, see Fig. (\ref{Dichtesigma40}). Close to the criticality, this behaviour is reminiscent of the well-known phenomenon of the critical slowing down near the point of a phase-transition. Critical slowing down goes along with the existence of long-ranged fluctuations of the systems internal modes. It may be these long-ranged fluctuations that stabilize the wave-packet against the influence of the attractive self-interaction above $\gamma_{c}$.

For attractive particle-interactions $\gamma\,<\,-0.627$, the internal motion of the wave-packet is not capable to stabilize the sytem. As the system evolves in time the wave-packet suffers a collapse. The collapse heats up the system because the interaction-energy is almost completely transformed into kinetic energy; along with the heating up, the system experiences a sudden loss of large fractions of particle density.

Our results clearly show, that the mechanism of quantum reflection remains a promising tool to trap cold atoms when particle interaction is included and tuned. However, as our variational analysis has revealed, the parameters of the system cannot be chosen arbitrarily. For best results, they must allow the system to evlove in the universal region close to $\gamma_{c}$. 

For alkali atoms, there are realistic surviving particle densities up to 50 \% for times around half a second. From table (\ref{couplingconstants}), the most promising candidate for trapping should be lithium, where the particle-interaction strength is already attractive.

\end{document}